\documentclass[twocolumn,showpacs,amsmath,floatfix,prb,aps]{revtex4}

\usepackage{graphicx}
\usepackage{dcolumn}
\usepackage{bm}

\newcommand{\bk}{{\bf k}}
\newcommand{\be}{\begin{equation}}
\newcommand{\ee}{\end{equation}}
\newcommand{\bea}{\begin{eqnarray}}
\newcommand{\eea}{\end{eqnarray}}
\newcommand{\mone}{{\bf 1}}
\newcommand{\br}{{\bf R}}
\newcommand{\bq}{{\bf q}}
\newcommand{\bs}{{\bf S}}

\begin{document}

\title{Canonical band theory of non-collinear magnetism}

\author{S. Shallcross$^{1}$}
\email{sam_shallcross@yahoo.co.uk}
\author{S. Sharma$^{2,3}$}
\affiliation{1 Department of Physics, Technical University of Denmark,\\
Building 307, DK-2800 Kgs. Lyngby.}
\affiliation{2  Fritz Haber Institute of the Max Planck Society, Faradayweg 4-6,
D-14195 Berlin-Dahlem, Germany.}
\affiliation{3 Institut f\"{u}r Theoretische Physik, Freie Universit\"at Berlin,
Arnimallee 14, D-14195 Berlin, Germany}

\date{\today}

\begin{abstract}

A canonical band theory of non-collinear magnetism is developed and applied
to the close packed fcc and bcc crystal structures. Several examples of non-collinear
magnetism in the periodic table are seen to be canonical in origin. This is a
parameter free theory where the crystal and magnetic symmetry, and exchange splitting,
uniquely determine the electronic bands. 
The only contribution to the determination of magnetic stability is the 
change in band energy due to hybridisation resulting from spin mixing, and on
this basis we are able to analyse the origin of the stability of non-collinear 
magnetic structures, and the instability of the FM state towards non-collinear
ordering.

\end{abstract}


\maketitle

\section{Introduction}

Structural trends in the periodic table are now very well understood, 
in contrast the situation for magnetic trends is not so clear. This
can be attributed both to the fact that only a subset of the periodic table
is magnetic in the three dimensional solid - the Fe group and rare earths -
and also the much greater degree of freedom for magnetic ordering as compared
to structural ordering. Indeed, within the transition metal block (with the
exception of Mn) all elements are found either as fcc, hcp, or
bcc lattices whereas a wealth of magnetic structures may be found.
One has has both longitudinal and helical spin density waves,
ferromagnetism, antiferromagnetism, as well as other more general non-collinear
structures.

For the Fe-group metals at ambient pressure the situation
is that near the centre of the series one typically finds antiferromagnetic (AFM)
structures whilst near the ends ferromagnetic (FM). In between one has
several metals - $\gamma$-Fe (fcc Fe), and bcc and fcc Mn - for
which non-collinear structures can be found \cite{kub00}. On the other hand, the late 
rare earth group elements are non-collinear, whilst in the middle of the series
one has ferromagnetic Gd \cite{jenmac91}. 

Of course, in a fully relativistic picture the
magnetisation vector field will always, to some extent, be non-collinear.
However, for the Fe-group and rare earths this leads only to intra-atomic
non-collinearity and not to the coarse grained inter-atomic 
non-collinearity where the average moments associated with different sites 
point in different directions.

The richness of magnetic structures means that most theoretical approaches
focus on specific materials, and only a few attempts have been
made to determine conditions whereby one places possible magnetic
structures in some general scheme. A notable early attempt along such lines
was the work of Pettifor \cite{pet80} who,
on a similar basis to Friedel's theory for the stability of crystal structures,
formulated a phase diagram in terms of
the Stoner parameter to the band width $I/W$ and the $d$-band 
occupation number.
In this theory the ferromagnetic and disordered local
moment structures were considered, and the phase diagram revealed a
simple argument as to why bcc Fe is a good
local moment system, but fcc Co and fcc Ni are not.
Heine and Samson \cite{hei83}, again using arguments based on
generalised Stoner criteria constructed a similar phase diagram diagram but
included also the AFM structure. They further
made actual calculations of the generalised Stoner $I$ within the tight
binding approximation, showing how $\gamma$-Fe
was placed at the crossing point of the AFM and FM stability criteria, and hence
was likely to assume a non-collinear structure.
Hirai, \cite{hir82} on the basis of an approximation to Hartree-Fock theory, calculated
the energy of FM, AFM, and helical spin spiral structues as a function
of an intra-atomic parameter and the $d$-band occupation number.
The resulting phase diagram showed the appearence of a region of helical
stability inbetween regions of FM and AFM stability. Unfortunately, the
electronic structure in the Hartree-Fock approximation is known to be
drastically in error compared to experiment.

Recently, an attempt was made to re-examine the issue of non-collinear stability 
on the basis of first principle calculations \cite{lars04}. Interestingly, 
it was found that materials collinear at their equlibrium moment were 
non-collinear for smaller (local) moments, induced either by pressure or a 
fixed spin moment procedure.
On this basis, it was speculated both that any magnetic material could be
made non-collinear and also that the primary factor governing the instability of the FM
state towards a non-collinear
state was the hybridisation of crossing spin-up and spin-down bands at the Fermi level.
However, in contrast to previous approaches no attempt was made to link this
to the $d$-band occupation.

The purpose of the present article is to unify the spirit of the earlier works,
where the emphasis was upon finding general magnetic phase diagrams,
with the approach in Ref.~\onlinecite{lars04} where the importance of spin hybridisation
for specific magnetic structures was emphased. 
As we shall show, this can be accomplished on the basis of
Andersen's canonical $d$-band theory \cite{and75,and77}, suitably generalised 
to deal with non-collinear magnetism. It has been shown in the past that
canonical $d$-band theory gives a good qualitative account of crystal
stability in the transition metal block and rare earths \cite{and77,skriv85}, 
even including the impact of collinear magnetism upon crystal
stability \cite{sod94,ab96}. Until now, however, it has not been used to 
investigate non-collinear magnetic stability.

The remainder of this article is structured as follows. In Section II we
describe canonical band theory and its generalisation to deal
with non-collinear magnetism, followed in Section III by details
pertinent to its implementation. In Section IV we present non-collinear
magnetic phase diagrams for the close packed lattices, and in Section V
discuss the reasons for non-collinear stability. Finally, in Section VI we
conclude.

\section{Non-collinear canonical band theory}

Within the Linear Muffin Tin Orbital
electronic structure method in the atomic sphere approximation (LMTO-ASA)
the electronic bands are given by the secular equation

\be
[P_{L'L}(E)\delta_{l'm'lm} - S_{L'L}(\bk)]a_{i\bk} = 0
\ee
where $L=(lm)$ are angular momentum indices, $\bk$ a reciprocal lattice
vector, and $i$ a band index.
This equation has the remarkable feature that it is split
into a part which depends solely on the one-electron potential for the given
atomic species, $P_{L'L}(E)$, known as the potential function, and a part
which depends only on the crystal symmetry,
the structure matrix $S_{L'L}(\bk)$.
Neglecting the off diagonal $l'l$ blocks of the structure matrix, which
amounts to neglecting all hybridisation effects between different angular
momentum chanels, leads to a set of pure $l$ canonical bands given by
$P_{li} = S_{li}(\bk)$ unique for each crystal structure.
In cases where the physics of the $d$-band is expected to play a leading
role one can further neglect all but the $l'=l=2$ block and by approximating the
potential function by $P=(E-C)/(\mu S^2)$ find a simple eigenvalue problem
for the canonical $d$-bands

\be
[(\epsilon_{i\bk} - C)/(\mu S^2) \mone - {\bf S}(\bk)]a_{i\bk} = 0
\ee
where $1 \le i \le 5$.
One should note that since $C$ is the $d$-band centre and $1/(\mu S^2)$ simply
sets the energy scale, this is a parameter free theory.
 
To develop a non-collinear magnetic generalisation one
first must specify a particular class of magnetic structures. It is
convenient to consider helical spin spiral structures, since these contain
both the FM and AFM structures as limits. In a helical spin spiral
structure the magnetisation is given by

\be \label{eq:ss}
m(\br_n) = (\sin \theta \cos (\bq.\br_n),\sin \theta \sin (\bq.\br_n),\cos \theta)
\ee
with $\br_n$ a direct lattice vector and $\bq$ a reciprocal lattice vector.
Although this structure may be incommensurate with the underlying lattice, it
is invarient under a lattice translation $\br_n$ and a rotation about the
spiral axis by $\bq.\br_n$. This leads to a generalised Bloch theorem
and an LMTO-ASA secular equation. By deploying the canonical procedure
outlined above one finds an eigenvalue problem for the canonical bands
of the spin spiral given by

\be \label{eq:ceig}
[(\epsilon_{i\bk\sigma} - C)/(\mu S^2) \mone - 
 ({\bf S}(\bk)+{\bf\Delta})]a_{i\bk\sigma} = 0
\ee
where

\begin{equation}
{\bf S}(\bk) = 
\left( \begin{array}{cc}
\bs^{+} + \bs^{-} \cos\theta &
\bs^{-} \sin\theta \\
\bs^{-} \sin\theta &
\bs^{+} - \bs^{-} \cos\theta
\end{array} \right) 
\end{equation}
and $S_{m'm}^{\pm} = S_{m'm}^{{\bf k}+{\bf q}/2} \pm S_{m'm}^{{\bf k}-{\bf q}/2}$, 
and also

\begin{equation}
{\bf \Delta} =
\left( \begin{array}{cc}
\Delta/2 &
0 \\
0 &
-\Delta/2
\end{array} \right) 
\end{equation}.
Eq.~\ref{eq:ceig} determines a set of bands unique for each symmetry of the
spin space group.

To complete the theory one needs to determine the canonical moment $m$
projected onto the local quantisation axis, for a given exchange splitting.
This may be extracted by evaluating the expectation value of $\sigma_z$
over the canonical wave function which leads to the natural result

\be \label{eq:cm}
m = \sum_{i\bk}^{occ}(|a_{i\bk\uparrow}|^2 - |a_{i\bk\downarrow}|^2)
\ee 
and where the sum is over all occupied states. Together with the $d$-band
occupation number
$n$ and the one-electron energy $e$, given respectively by

\bea \label{eq:cne}
n & = & \sum_{i\bk\sigma}^{occ}|a_{i\bk\uparrow}|^2 \nonumber \\
e & = & \sum_{i\bk\sigma}^{occ}|a_{i\bk\sigma}|^2 \epsilon_{i\bk\sigma}
\eea
these form the basic quantities of the theory. However, in order to determine
the stability of competing magnetic structures for a given $n$ and $m$ what
one needs is not the one-electron energy but the kinetic energy. This
is due to the fact that in a canonical theory there is no intrinsic intra-atomic
force causing the magnetisation, but rather it is imposed by setting the
exchange splitting to some finite value which thus plays the role of an 
external field. The contribution of the energy of this external field
is included in the one-electron energy and so must be subtracted, which
leaves the kinetic energy of the one-electron system

\be \label{eq:cT}
T = e - \int_0^m \Delta(m) dm
\ee
as the quantity to be used for comparing the stability of different magnetic
structures.

\section{Computational details}

In order to solve Eqs.~\ref{eq:ceig} to \ref{eq:cne} we use a Monkhurst-Pack mesh
with 18900, 109800, 216000, and 37820 $\bk$-points in the irreducable Brillouin zone for
the FM structure, and spirals along $\Gamma X$, XW and WL, and LG respectively.
For the bcc irreducable Brillouin zone we use
5200, 28830, and 109800 $\bk$-points for the FM structure and spirals
along $\Gamma$H, and HP respectively.
For the $\bk$-integration in Eqs.~\ref{eq:cm} and \ref{eq:cne} we
use an eigenvalue broadening of 4.5mRy.
For the potential parameters which, for the canonical $d$-band theory, simply
set the origin and scale of energy we set $C=0$ and take $1/(\mu S^2)$ for
non-magnetic fcc Fe.

One should note that the natural variables of the theory outlined above 
are the Fermi energy
$E_F$ and the exchange splitting $\Delta$.
However, the physical
variables for determining magnetic stability are the $d$-band occupation
number $n$ and the spiral local moment $m$. Thus we need to solve
$n'(E_F,\Delta,\bq)=n$ and $m'(E_F,\Delta,\bq)=m$ to find the values of
$E_F$ and $\Delta$ that give the required $n$ and $m$. This can easily be
done by minimising a function defined as

\be \label{eq:cmin}
f = [n'(E_F,\Delta,\bq)-n]^2 + [m'(E_F,\Delta,\bq)-m]^2.
\ee
In order to facilitate this it is convenient to pre-calculate the
functions $n'$, $m'$, and also the corresponding one-electron energy function
$e'(E_F,\Delta,\bq)$ over the full range of $E_F$ and $\Delta$ and interpolate
using cubic splines. In this way one can very rapidly minimise Eq.~\ref{eq:cmin},
whereas if $n'$ and $m'$ are repeatedly calculated by solving Eq.~\ref{eq:ceig}
over the Brillouin zone this would be a very slow procedure.

\section{Canonical phase diagrams}

\begin{figure}
\caption{Canonical phase diagram for $\Gamma XW$ spin spirals in the fcc structure.
}
{\includegraphics[scale=0.8,angle=00]{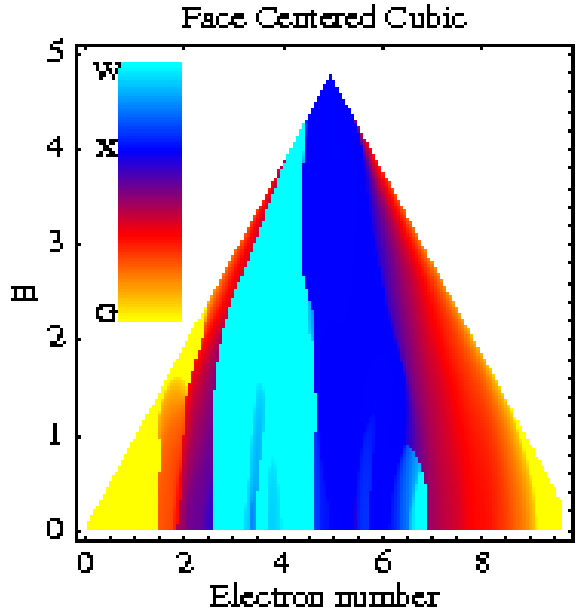}}
\label{fig:pdGXW}
\end{figure}

\begin{figure}
\caption{Canonical phase diagram for $\Gamma HP$ spin spirals in the bcc structure.
}
{\includegraphics[scale=0.8,angle=00]{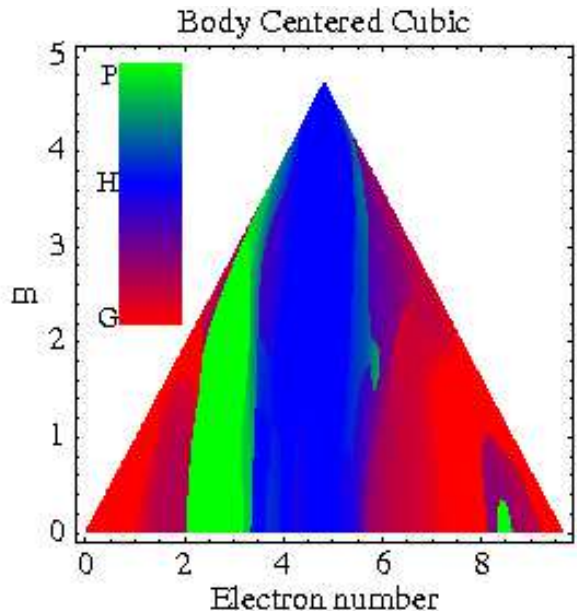}}
\label{fig:pdGHP}
\end{figure}

In Figs.~\ref{fig:pdGXW} and \ref{fig:pdGHP} we show the calculated phase
diagrams when one considers spin spirals along the $\Gamma XW$ symmetry
lines of the fcc Brillouin zone and the $\Gamma HP$ lines of the bcc Brillouin zone.
In order to construct these phase diagrams we have used planer spin spin spirals,
$\theta=\pi/2$ in Eq.~\ref{eq:ss}, since then AFM structures are limits
of spin spirals. In fact, for single basis crystal structures like fcc and bcc,
all points of the BZ boundary correspond to AFM structures.

One can notice immediately that, as expected, near half filling of the $d$ band
one finds AFM structures while for the nearly fully occupied or empty $d$-band one finds
ferromagnetism. Unfortunately, the correspondence with
the $d$ electron numbers of the transition metals 
is only approximate, since the region of FM is too small in
both the fcc and bcc phase diagrams. For instance, in the fcc phase diagram FM is
the stable solution for $n > 8.8$, whereas since fcc Co is FM the boundary should
be $n \approx 8.0$.

\begin{figure}
\caption{Canonical spin spiral spectrum's for the symmetry path $\Gamma$XWL$\Gamma$ in the
fcc Brillouin zone. Shown are spectrums for (a) $n=9.3$, $m=0.5$, (b) $n=8.8$, $m=0.4$, (c)
$n=7.5$, $m=1.2$, and (d) $n=0.5$, $m=3.3$. Shown in the lowest panel is (e) the region
around the X point for $n=6.5$, $m=0.8$. The top scale refers to graphs (a)-(d), and the
bottom to graph (e).
\vspace{0.7cm}
}
{\includegraphics[scale=0.3,angle=00]{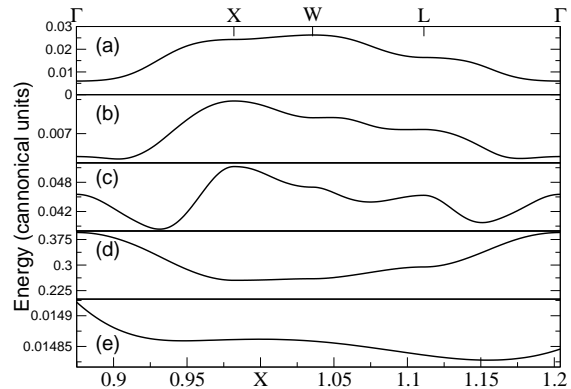}}
\label{fig:fspec}
\end{figure}

In Ref.~\onlinecite{lars04} it was noted that fcc Co developed a spin
spiral structure along the $\Gamma X$ direction if the moment was reduced from its
equlibrium value to 0.8 $\mu_B$. In fact, as one can see from Fig.~\ref{fig:pdGXW}
this is what happens in the canonical picture as well: there is a transition from
FM stability to a spin spiral in the $\Gamma X$ direction as the moment is reduced
for $n \approx 8.8$. Interestingly, reducing the value of $n$ continuously
increases this splitting about the $\Gamma$ point so that it becomes a spiral
with $\bq = [0,0,0.5]$, which is exactly the spin spiral structure observed
for $\gamma$-Fe in calculations. In fact, in the fcc phase diagram it can
be seen that there is quite a large region where this spiral is stabilised.
Furthermore, by lowering the moment
for $n \approx 6.5$ this spiral closes about the $X$ point until one finds
nearly degenerate spin spirals along the $\Gamma X$ and $XW$ symmetry lines,
again as is seen in self-consistent calculations. Thus one can see that the
formation of small moment spin spirals for larger $n$ moves continuously
to the intermediate moment spin spiral of $\gamma$-Fe before finally the $X$
AFM structure becomes stable near half filling. This is an appealing
picture, since it unites the results of Ref.~\onlinecite{lars04} into a general
magnetic phase diagram for the fcc structure. In Fig.~\ref{fig:fspec} are shown
the spin spiral spectrum's corresponding to the various magnetic phases discussed.

For the bcc structure, shown in Fig.~\ref{fig:pdGHP} one can note that
the region of stability for the FM structure is much larger, which makes
sense since bcc Fe is ferromagnetic whereas $\gamma$-Fe is non-collinear.
However, as for the fcc lattice the region of FM stability is too small.
Interestingly, at $n \approx 9.0$ one finds
an incursion of non-collinear structures for small moments.
Again, this is what is seen in Ref.~\onlinecite{lars04}, where bcc Ni 
becomes non-collinear for small moments.

Finally, one can note the approximate symmetry of both the fcc and bcc 
magnetic phase diagrams about half filling. This is interesting since,
as was already mentioned, the late rare earths all assume a non-collinear
structure. Since for these metals the $f$ electrons are localised and
simply play the role of polarising the itinerent $spd$ electrons one
might expect that again the crucial parameters for describing their
magnetism would be the $d$-band electron number and moment $m$. In
fact, this is exactly what happens for the structural energies where
for $1.5 \le n \le 2.0$ one finds the sequence of rare earth structures
\cite{skriv85}. The symmetry seen in Figs.~\ref{fig:pdGXW} and \ref{fig:pdGHP}
is suggestive of the fact that there is a connection between the
late rare earths and the late Fe-group elements with artifically lowered 
moments. However, in contrast to the canonical crystal energy differences,
the magnetic canonical theory is not sufficiently quantitative to
give the actual non-collinear structures of the rare earths, and therefore
we have not shown the hcp phase diagram.

For a more quantitative theory,
the effect of angular momentum hybridisation must be included, but in this
case it is no longer a parameter free theory, since band centres, band
distortions, and band widths must be introduced for the $l=0,1,2$ chanels.

\section{Origin of stability of non-collinear structures}

\begin{figure}
\caption{Canonical Density of States for the (a) ferromagnetic, (b) $\bq = 
[0.0,0.0,0.5]$ spin spiral, and (c) antiferromagnetic X-point structure.
\vspace{1.0cm}
}
{\includegraphics[scale=0.3,angle=00]{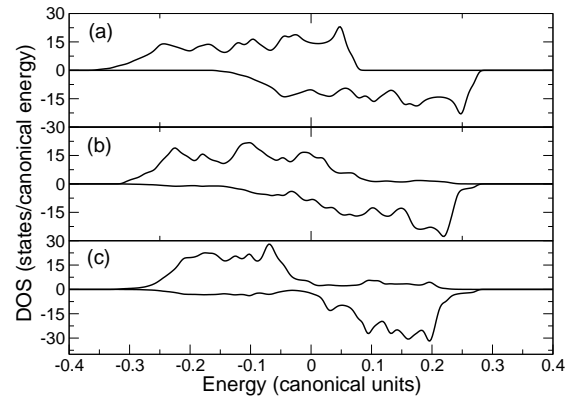}}
\label{fig:dos}
\end{figure}

It is interesting to first reconsider the arguments for the stability
of the collinear AFM and FM structures. The reason for the existance
of AFM near half filling and FM for the nearly full band is well
known and has been described by a number of authors (see Ref.~\onlinecite{kub00}
and references therein). The hybridisation between the different sublattices
of the AFM structure, equivalent to the hybridisation from
spin mixing when the AFM structure is considered as the limit of a planer
spin spiral, opens a large pseudo-gap in the density of states. It is this which
stabilises the AFM structure at half filling. The argument is thus couched
in terms of significant movements of spectral weight at all energies, and 
not to changes at only the Fermi level. In Fig.~\ref{fig:dos} are shown the
density of states (DOS) for the FM, X point AFM, and spin spiral $\bq = [0.0,0.0,0.5]$ 
structures for the fcc lattice. As can be seen the structure of the
spiral DOS is somewhere between that
for the FM and AFM structures.
Thus the occurance of non-collinear stability
between the FM and AFM regions can be understood by the same reasoning that
explains the placement of the FM and AFM regions themselves.

It is natural that the determination of the lowest energy magnetic structure
will, in general, come about as the result of movements of spectral weight at 
all energies since generally spin hybridisation will be very significant.
On the other hand, it is not immediately clear that this will be the case when
one considers the instability of the FM state towards non-collinear structures.
In Ref.~\onlinecite{lars04} it was speculated that such instabilities
are driven by the hybridisation
of crossing spin-up and spin-down bands at the Fermi level. This is, of course, the 
classic picture when one has the situation of Fermi surface nesting, as
happens in the late rare earths. There it is the Fermi surface of
the non-magnetic state which shows nesting and this is relevent to the
final magnetic state since the itinerant $spd$ moment is very 
small (0.2-0.6 $\mu_B$). However, the argument presented in Ref.~\onlinecite{lars04}
is that the number of spin-up and spin-down band crossings at the Fermi energy
may more generally hold the key to understanding the instability of the 
ferromagnetic state towards a non-collinear state.

The stability or otherwise of the FM state is given by the eigenvalues of the
Hessian of the FM
point in a general energy space of magnetic structures. For the magnetic structures
considered in this work,
this could be determined either as a $\theta \rightarrow 0$ limit for a finite
$\bq$ or alternatively as a $\bq \rightarrow [0,0,0]$ limit for finite $\theta$.
To investigate the origin of the instabilty of the FM state towards non-collinearity
we choose to investigate the energy difference of the FM structure with the 
$\bq = [0.0,0.0,0.1]$ spin spiral. This will be positive if the FM structure is
stable, and negative otherwise. If the arguments of Ref.~\onlinecite{lars04}
are relevent then this energy difference should be dominated by the contribution
from a window near the Fermi energy.

\begin{figure}
\caption{Energy difference of FM and $\bq = [0.0,0.0,0.1]$ spin spiral structures. Full
line is total energy different whilst caption indicates components.
\vspace{1cm}
}
{\includegraphics[scale=0.3,angle=00]{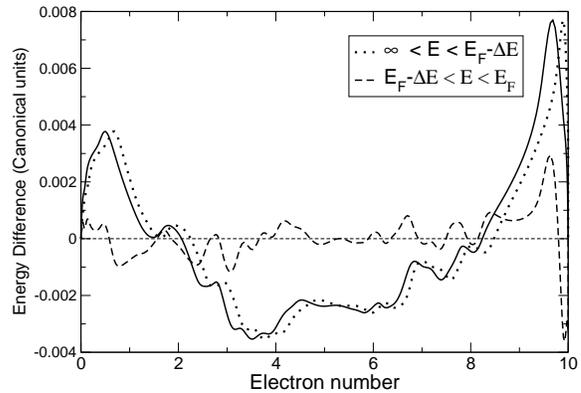}}
\label{fig:ediff}
\end{figure}

In Fig.~\ref{fig:ediff} is plotted the energy difference of the FM
and $\bq = [0.0,0.0,0.1]$ structures for the fcc lattice
as a function of band filling for a fixed exchange splitting
of $\Delta = 0.3$. This leads to a path in the (n,m) phase diagram
with $m$ almost at the saturation value for each $n$
(at half filling $m=4.2$).
Also shown in Fig.~\ref{fig:ediff} is a breakdown of the energy into that
arising from a small window near the Fermi energy, with $\Delta E =$ 0.015, 
and that from the remaining spectrum.
One can clearly see that it is this energy and not that arising from the Fermi 
energy window which follows the total energy difference.
In fact, while the indication of FM stability from the total energy difference is
in agreement with what one would expect from the phase diagram in Fig.~\ref{fig:pdGXW},
the Fermi window energy difference shows no correspondence to it.
One can also notice that the energy difference has exactly two
crossing points. This is in line with the theorem of Heine and Samson \cite{hei80}
stating that any property of a tight binding band must have at least two crossing
points as a function of band filling. A similar picture is found for other
settings of the exchange splitting.

It would appear then that there is no special significance attached to
spin hybridisation at the Fermi level in considering the stability of the
FM state.
This makes sense when one considers that, for a given band filling, if
the hybridisation arising from solely around the Fermi level 
dominated, then the corresponding change for some lower energies must
integrate to zero. This could only be the case if the change in band structure from
hybridisation is exactly symmetric. This cannot, of course, be the case for
all energies i.e. for all $d$-band electron numbers. Thus, in general,
it is hybridisation changes for all energies that determine even the 
stability criteria of the FM state towards non-collinearity.

\section{Conclusions}

To conclude a generalisation of Andersen's canonical band theory has been
used to derive magnetic phase diagrams for the close packed bcc and fcc
structures. One finds that a number of the non-collinear structures
observed in self-consistent calculations are found, but that the phase
diagrams are in error for moments near saturation and the regions of ferromagnetic
stability are too small. This is most likely due to the missing
hybridisation between different angular momentum chanels. A similar situation arises
in the case of crystal stability where errors found in the $d$-band canonical 
theory are removed by inclusion of $s$ and $p$ states \cite{skriv85}.

An analysis of the canonical magnetisation energetics in terms of 
contributions from different energy windows,
shows that, generally, it is hybridisation from the whole
spectrum which acts to determine the stability of the ferromagnetic state
with regard to non-colliear magnetism.
Thus the picture of band crossing at the Fermi level
emphased in Ref.~\onlinecite{lars04} does not appear to be relevant
for the determination of the stability of the ferromagnetic state.

\begin{acknowledgments}

S. Shallcross would like to acknowledge fruitful converstions with 
Lars Nordstr\"om and especially stimulating conversations with Hans L. Skriver.
S. Sharma would like acknowledge NoE NANOQUANTA Network (NMP4-CT-2004-50019) 
and Deutsche Forschungsgemeinschaft for financial support.

\end{acknowledgments}

\end{document}